\documentclass[amsmath,amssymb,aps,prl,twocolumn,superscriptaddress,floatfix,nopacs,preprintnumbers, nolongbibliography,nofootinbib] {revtex4-2}

\usepackage[utf8]{inputenc}
\usepackage{subfigure}
\usepackage[normalem]{ulem}
\usepackage[dvipsnames,table]{xcolor}
\usepackage{mathrsfs}
\usepackage{amssymb,bm}
\usepackage{dsfont}
\usepackage{amsmath}
\usepackage{mathtools}
\usepackage{slashed}
\usepackage{xfrac}
\usepackage{graphics}
\usepackage{graphicx}
\usepackage{physics}
\usepackage[h]{esvect}
\usepackage{accents}
\usepackage{standalone} 
\usepackage{graphicx}
\usepackage{tikz}
\usepackage{verbatim}
\usepackage{ulem}
\standaloneconfig{mode=buildnew}

\definecolor{lcolor}{rgb}{0.5,0,0}
\definecolor{citcolor}{rgb}{0,0.3,0.0}

\renewcommand{\dd}{\mathrm{d}}

\usepackage{xargs}
\usepackage[colorinlistoftodos,prependcaption,textsize=footnotesize]{todonotes}
\usepackage[breaklinks,colorlinks,citecolor=citcolor,urlcolor=blue,linkcolor=lcolor]{hyperref}

\begin{document}

\preprint{RIKEN-iTHEMS-Report-26}

\author{Gert Aarts}
\email{g.aarts@swansea.ac.uk}
\affiliation{Centre for Quantum Fields and Gravity, Department of Physics, Swansea University, Swansea SA2 8PP, United Kingdom}

\author{Diaa E.~Habibi}
\email{n.e.habibi@swansea.ac.uk}
\affiliation{Centre for Quantum Fields and Gravity, Department of Physics, Swansea University, Swansea SA2 8PP, United Kingdom}

\author{Andreas Ipp}
\email{ipp@hep.itp.tuwien.ac.at}
\affiliation{Institute for Theoretical Physics, TU Wien, Wiedner Hauptstraße  8-10/136, A-1040 Vienna, Austria}

\author{David I.~M\"{u}ller}
\email{dmueller@hep.itp.tuwien.ac.at}
\affiliation{Institute for Theoretical Physics, TU Wien, Wiedner Hauptstraße  8-10/136, A-1040 Vienna, Austria}

\author{Thomas R.~Ranner}
\email{thomas.ranner@tuwien.ac.at}
\affiliation{Institute for Theoretical Physics, TU Wien, Wiedner Hauptstraße  8-10/136, A-1040 Vienna, Austria}

\author{Lingxiao Wang}
\email{lingxiao.wang@riken.jp}
\affiliation{RIKEN Center for Interdisciplinary Theoretical and Mathematical Sciences (iTHEMS), Wako, Saitama 351-0198, Japan}
\affiliation{Institute for Physics of Intelligence, Graduate School of Science, The University of Tokyo, Bunkyo-ku, Tokyo 113-0033, Japan}

\author{Wei Wang}
\email{wei.wang@sjtu.edu.cn}
\affiliation{State Key Laboratory of Dark Matter Physics, Key Laboratory for Particle Astrophysics and Cosmology (MOE),
Shanghai Key Laboratory for Particle Physics and Cosmology,
Shanghai Jiao Tong University, Shanghai 200240, China}
\affiliation{Southern Center for Nuclear-Science Theory (SCNT), Institute of Modern Physics, Chinese Academy of Sciences, 96 South Sihuan Rd. Huicheng District, Huizhou 516000, Guangdong, China}

\author{Qianteng Zhu}
\email{zhuqianteng@sjtu.edu.cn}
\affiliation{RIKEN Center for Interdisciplinary Theoretical and Mathematical Sciences (iTHEMS), Wako, Saitama 351-0198, Japan}
\affiliation{State Key Laboratory of Dark Matter Physics, Key Laboratory for Particle Astrophysics and Cosmology (MOE),
Shanghai Key Laboratory for Particle Physics and Cosmology,
Shanghai Jiao Tong University, Shanghai 200240, China}

\title{Generalizable Equivariant Diffusion Models for Non-Abelian Lattice Gauge Theory }

\begin{abstract}
We demonstrate that gauge equivariant diffusion models can accurately model the physics of non-Abelian lattice gauge theory using the Metropolis-adjusted annealed Langevin algorithm (MAALA), as exemplified by computations in two-dimensional U(2) and SU(2) gauge theories. Our network architecture is based on lattice gauge equivariant convolutional neural networks \mbox{(L-CNNs)}, which respect local and global symmetries on the lattice. Models are trained on a single ensemble generated using a traditional Monte Carlo method. By studying Wilson loops of various size as well as the topological susceptibility, we find that the diffusion approach generalizes remarkably well to larger inverse couplings and lattice sizes with negligible loss of accuracy while retaining moderately high acceptance rates.
\end{abstract}

\maketitle 

\textit{Introduction.}
Lattice quantum chromodynamics (QCD) is an invaluable computational tool to investigate the non-perturbative aspects of the strong nuclear force, which describes the dynamics of quarks and gluons \cite{FlavourLatticeAveragingGroupFLAG:2024oxs,Aarts:2023vsf}. In this approach to quantum field theory, matter and gluon fields are defined on a four-dimensional space-time lattice to enable the numerical computation of physical observables using Monte Carlo (MC) integration. To make accurate physical predictions free from lattice artifacts, it is necessary to perform simulations at successively decreasing lattice spacing by adjusting the inverse coupling constant $\beta$. Continuum extrapolations are associated with exponentially increasing computational cost due to \textit{critical slowing down} \cite{Schaefer:2010hu}. At large $\beta$, simulations may also suffer from \textit{topological freezing}, where the phase space is inefficiently explored due to strongly suppressed tunneling between topological sectors \cite{Alles:1996vn,DelDebbio:2004xh}. Many alternative computational methods have been proposed, such as parallel tempering \cite{Hasenbusch:2017unr, Bonanno:2020hht, Bonanno:2024zyn}, flow-based approaches (normalizing \cite{Kanwar:2020xzo,Boyda:2020hsi,Abbott:2024kfc}, continuous or trivializing \cite{deHaan:2021erb, DelDebbio:2021qwf, Gerdes:2022eve, Bacchio:2022vje, Albandea:2023wgd, Gerdes:2024rjk}), stochastic normalizing flows \cite{Caselle:2022acb, Bulgarelli:2024brv, Bonanno:2025pdp}, and, very recently, machine learned parametrizations \cite{Holland:2024muu, Holland:2025fsa} of fixed-point actions with strongly suppressed lattice artifacts \cite{Hasenfratz:1993sp}.

\begin{figure}[!tbp]
    \centering
    \includegraphics[width=.95\linewidth]{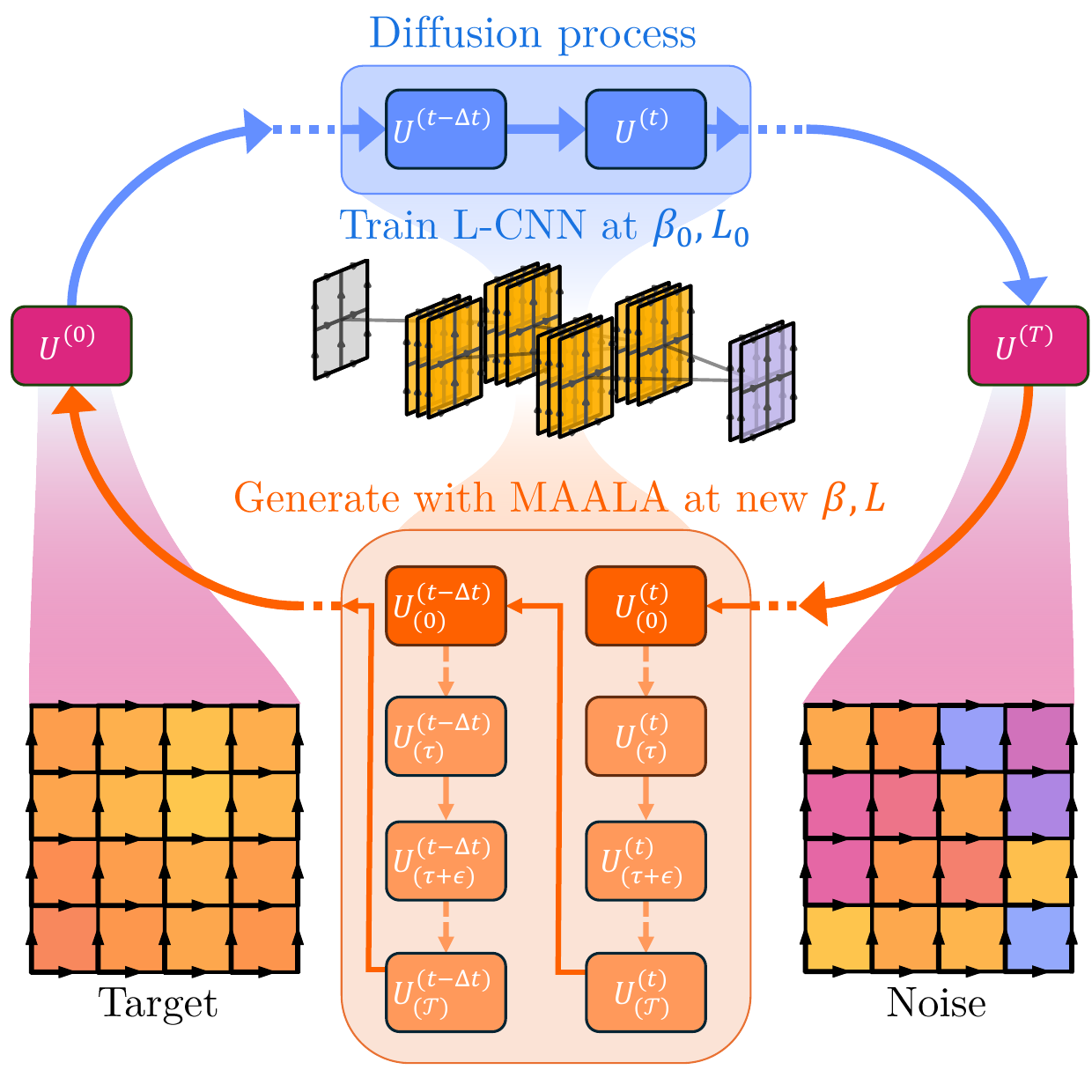}
    \caption{Schematic overview of the diffusion and generative process using a lattice of gauge links (edges) forming plaquettes (colorful faces). Training samples from the target distribution at time $t=0$ undergo diffusion until they resemble uncorrelated noise at later time $T$. We train an L-CNN with diffused samples at a given inverse coupling $\beta_0$ and lattice size $L_0$. After training, we solve the reversed diffusion process using the Metropolis-adjusted annealed Langevin algorithm (MAALA) to generate new independent samples at arbitrary $\beta$ and $L$.}
    \label{fig:denoise}
\end{figure}

As part of a broad program to incorporate machine learning approaches in (lattice) QCD \cite{Cranmer:2023xbe,Aarts:2025gyp,Tomiya:2025quf}, we discuss here a new approach to address these issues, namely diffusion models (DMs), a class of generative machine learning models that have gained popularity through their state-of-the-art performance in image generation \cite{sohldickstein2015deepunsupervisedlearningusing, song2020generativemodelingestimatinggradients, ho2020denoisingdiffusionprobabilisticmodels, song2021scorebasedgenerativemodelingstochastic}. Given a set of training samples, DMs learn the evolution of a data distribution under a diffusive process towards a trivial distribution~\cite{Lai2025dm}. Once trained, this information allows for the construction of a reverse process, which can generate new, statistically uncorrelated samples by \textit{denoising} uncorrelated noise (see Fig.~\ref{fig:denoise}). Thus, DMs, like flow-based approaches, provide a potential solution to critical slowing down and topological freezing. There is a strong correspondence between continuous normalizing flows \cite{Gerdes:2022eve, Cheng:2025khv} and DMs, as each DM can be re-formulated as a deterministic flow. Moreover, there exist intriguing connections to stochastic quantization \cite{Wang:2023exq, Fukushima:2024oij} and the stochastic renormalization group \cite{Carosso:2019qpb, Cotler:2023lem,Masuki2025DMRG}. Motivated by lattice QCD, DMs have so far been applied to $\phi^4$-theory \cite{Wang:2023sry, Wang:2023exq}, U(1) gauge theory \cite{Zhu:2024kiu, Zhu:2025pmw, Vega:2025hgz} and in the context of complex actions \cite{Aarts:2025lpi}. 

In this Letter, we demonstrate that score-based DMs \cite{song2021scorebasedgenerativemodelingstochastic} can effectively sample non-Abelian lattice gauge theories---including out-of-domain sampling by extrapolating far away from the original training data in terms of coupling and lattice size---with high accuracy. This is achieved by combining lattice gauge equivariant convolutional neural networks (L-CNNs) \cite{Favoni:2020reg}, to model the score function, and the recently-developed Metropolis-adjusted annealed Langevin algorithm (MAALA) with score rescaling for the reverse process \cite{Zhu:2024kiu, Zhu:2025pmw}. We apply our model to two-dimensional U($2$) and SU($2$) lattice gauge theories and find that both local and global observables---Wilson loops of arbitrary size and the topological susceptibility---are accurately reproduced. This is the first demonstration that DMs can be used in non-Abelian gauge theories to generalize to gauge couplings and lattice sizes far beyond the provided training data, in a gauge equivariant manner. 

\textit{Monte Carlo simulations of lattice gauge theory}
provide a way to compute the expectation values of observables formulated as functional integrals,
\begin{equation} \label{eq:ltg_integral}
\langle O \rangle \! = \! \frac{1}{Z} \! \int \! \mathcal{D}\mathbf{U} \,O(\mathbf{U}) \, e^{-S_E(\mathbf{U})}, \quad Z \! = \! \! \int \! \mathcal{D}\mathbf{U} \, e^{-S_E(\mathbf{U})},
\end{equation}
where $Z$ is the partition function and $S_E(\mathbf{U})$ is the standard Wilson action
\begin{equation}
\label{eq:wilson_action}
S_E(\mathbf{U}) = \frac{\beta}{N_c} \sum_{x, \mu < \nu} \mathrm{Re} \, \mathrm{Tr} \,  [\mathds{1} - P_{x, \mu\nu}].
\end{equation}
Here, $\beta = 2 N_c / g_\mathrm{YM}^2 $ with $g_\mathrm{YM}$ the Yang-Mills coupling and $N_c \geq 1$ specifies the gauge group $G$, either SU($N_c$) or U($N_c$). The sum runs over the sites $x$ of a periodic $L^D$ lattice with $D \geq 2$ dimensions, and the indices $0 \leq \mu, \nu < D$. The fields $\mathbf U$ consist of link variables $U_{x,\mu} \in G$. Plaquettes are given by $P_{x, \mu\nu} \equiv U_{x,\mu} U_{x+\hat{\mu},\nu} U_{x+\hat{\nu},\mu}^\dagger U_{x,\nu}^\dagger$ which span a square loop in the $(\mu\nu)$-plane.  

A defining feature of lattice gauge theories is that gauge invariance is exactly preserved at the discrete level, mirroring the fundamental symmetry of the continuum theory. Links and plaquettes transform as
\begin{equation} \label{eq:gauge_transformation}
    U'_{x,\mu} = \Omega_x U_{x,\mu} \Omega^\dagger_{x+ \hat \mu}, \qquad  P'_{x,\mu\nu} = \Omega_x P_{x,\mu\nu} \Omega^\dagger_{x},
\end{equation}
where $\Omega_x \in G$. Under these transformations, the Wilson action remains invariant, $S_E(\mathbf{U}') = S_E(\mathbf{U})$.

Direct computation of path integrals is often intractable, particularly in higher dimensions and non-perturbative regimes. Traditional Markov Chain Monte Carlo (MCMC) simulations compute $\langle O \rangle$ as a Monte Carlo integral by sampling $\mathbf U$ from the probability distribution $p(\mathbf{U}) \propto \exp[-S_E(\mathbf{U})]$. The computational challenge is to generate a large number of statistically independent samples. Particularly in the physically relevant regimes at large $\beta$ and large lattice volumes $V=L^D$, correlations become significant (critical slowing down) and the computational cost grows quickly.

\textit{Gauge equivariant diffusion models.}
Score-based DMs employing stochastic differential equations (SDEs) define the forward diffusion process as a type of Brownian motion. DMs in flat space have been covered extensively \cite{song2021scorebasedgenerativemodelingstochastic}. For applications to lattice gauge theory, the diffusion process must be adapted to the geometry of the group manifold (cf.~\cite{Batrouni:1985jn,Damgaard:1987rr} in the context of stochastic quantization and \cite{lou2023scaling, Kanwar:2025wuc} for DMs) and be compatible with gauge symmetry. Here, we introduce a fully gauge equivariant approach to DMs.

We define the group-preserving forward process as the solution to the Stratonovich SDE\footnote{Unlike the Itô prescription, the Stratonovich stochastic integral preserves the chain rule, which makes it suitable for DMs on Riemannian manifolds \cite{Damgaard:1987rr,huang2022riemannian}.}
\begin{align} \label{eq:forward_gauge}
    \!\!  \dd U^{(t)}_{x,\mu} = i \! \sum_a T^a \!\!\left[ K^a_{x,\mu}(\mathbf U^{(t)}) \dd t + g(t) \dd W^{(t), a}_{x,\mu}\right] \! \circ  U^{(t)}_{x,\mu},
\end{align}
where $T^a$ are the generators of the group and $a$ are color indices, $1 \leq a \leq D_A$ with $D_A = N_c^2 - 1$ for SU($N_c$) or $N_c^2$ for U($N_c$). The diffusion time $t$ is limited to $0 \leq t \leq T$, $K^a_{x,\mu}$ defines the drift, $g(t)$ is the diffusion coefficient, and $W^{(t), a}_{x,\mu}$ are independent Wiener processes. 
Adapting the analysis on flat space \cite{anderson1982reverse} to the case here,
the reverse process corresponding to Eq.~\eqref{eq:forward_gauge} is given by
\begin{align}\label{eq:reverse_gauge}
     \dd U^{(t)}_{x,\mu} &= i \sum_a T^a \big[ \big( K^a_{x,\mu}(\mathbf U^{(t)})- g(t)^2 D^a_{x,\mu} \ln p_t (\mathbf U^{(t)}) \big) \dd t \nonumber \\
    &\qquad\quad + g(t) \dd  \tilde W^{(t), a}_{x,\mu} \big] \circ U^{(t)}_{x,\mu},
\end{align}
with $t$ running backward from $T$ to $0$, $\tilde W^{(t), a}_{x,\mu}$ are reverse Wiener processes, and $D^a_{x,\mu}$ denotes the group derivative (see \hyperref[sec:endmatter]{End Matter}). For the reverse process, initial conditions are chosen at \mbox{$t=T$}, sampled from a prior distribution, $\mathbf U^{(T)} \sim p_T(\mathbf U)$. As we will see later, $p_T$ is the uniform Haar measure. The time-dependent probability distribution $p_t( \mathbf U)$ enters the reverse process through the \textit{score function} (the gradient of the log-likelihood function) $D^a_{x,\mu} \ln p_t(\mathbf U)$; both are generally unknown.

The diffusion processes are closely related to stochastic quantization (SQ) \cite{Parisi:1980ys,Damgaard:1987rr,Namiki:1993fd}. If the drift is chosen as \mbox{$K^a_{x,\mu}(\mathbf U) = - D^a_{x,\mu} S_E(\mathbf U)$} and $g(t) = \sqrt 2$, then for $t \rightarrow \infty$ the stochastic process in Eq.~\eqref{eq:forward_gauge} approaches a stationary distribution akin to the functional integral in Eq.~\eqref{eq:ltg_integral}. Moreover, the SQ process is compatible with gauge symmetry because the drift is a group derivative of a gauge invariant function $S_E(\mathbf U)$. Under gauge transformations \eqref{eq:gauge_transformation}, it then holds that
\begin{align}
    \hat K_{x,\mu}(\mathbf U') &= \Omega_{x}  \hat K_{x,\mu}(\mathbf U) \Omega^\dagger_{x}, \label{eq:drift_gauge_equiv}
\end{align}
where $\hat K_{x,\mu} = \sum_a T^a K^a_{x,\mu}$. As a result, the time-dependent probability distribution becomes invariant, $p_t(\mathbf U')  = p_t(\mathbf U)$, provided that the initial conditions at $t=0$ are invariant as well. 
If one requires Eq.~\eqref{eq:drift_gauge_equiv}, the gauge invariance property also holds for the DM processes in Eqs.~\eqref{eq:forward_gauge} and \eqref{eq:reverse_gauge}. 

The goal of score-based DMs is to approximate the score function by a neural network ${s}^a_{\theta;x,\mu}(\mathbf U, t)$. In our approach, largely following the denoising score-matching procedure \cite{6795935}, this is accomplished by minimizing the loss function
\begin{equation}
    \label{eq:gauge_loss_function}
    \int_0^T  \!\! {\dd t}  \!\!\!\!\!\!\!
    \underset{ \substack{ p_0(\mathbf{U}^{(0)}) \\ p_{0t}(\mathbf{U}^{(t)}|\mathbf{U}^{(0)}) }}
    {\mathbb{E}} \!\!
       \sum_{x,\mu,a} \!\!\big[ s_{\theta;x,\mu}^a(\mathbf{U}^{(t)} \!\!, t) \! - \! D^a_{x,\mu} \! \ln p_{0t}(\mathbf U^{(t)} | \mathbf{U}^{(0)}) \big]^2
\end{equation}
with respect to the network parameters $\theta$. The expectation value $\mathbb{E}$ is taken over the joint distribution of the training target $p_0(\mathbf{U}) \propto \exp[-S_E(\mathbf{U})]$ and the transition kernel  $p_{0t}(\mathbf{U}^{(t)}|\mathbf{U}^{(0)})$ as defined by the forward process in Eq.~\eqref{eq:forward_gauge}. The joint distribution in Eq.~\eqref{eq:gauge_loss_function} ensures that the network $s^a_{\theta;x,\mu}$ learns the score of the marginal distribution $p_t(\mathbf{U}^{(t)})$. To retain gauge invariance, the network must satisfy an analogous condition as in Eq.~\eqref{eq:drift_gauge_equiv}. With these constraints, the specific combination of terms in the loss function guarantees that the entire training process is compatible with local symmetry inherent to the physical theory.

Minimizing the loss function of Eq.~\eqref{eq:gauge_loss_function}, we are faced with the practical problem of numerically evaluating  $D^a_{x,\mu} p_{0t}(\mathbf{U}^{(t)} | \mathbf{U}^{(0)})$. 
Even for standard Brownian motion on compact groups, where $p_{0t}$ is given by the heat kernel, a simple closed-form expression is not available. Similarly, the solution of the forward SDE \eqref{eq:forward_gauge} must, in general, be obtained numerically. To avoid these issues which might impede efficient training, we instead define a new diffusion process, with $K^a_{x,\mu}=0$,
\begin{align} \label{eq:diffused_link}
    {U}^{(t)}_{x,\mu}  = \exp \big[ i \sigma(t)\sum_a T^a  \mathbf{\eta}^a_{x,\mu}  \big] {U}^{(0)}_{x,\mu}, 
\end{align}
where $\eta^a_{x,\mu}\sim {\mathcal N}(0,1)$ are independent Gaussian variables and  the noise scale is set to \mbox{$\sigma(t) = \sqrt{(s^{2 t} - 1) / (2 \ln s)}$} with $s=25$, i.e.\ a \textit{variance-expanding} scheme  \cite{song2021scorebasedgenerativemodelingstochastic}. At late times $t \approx T$, $\sigma(t)$ becomes large and the links are almost uniformly distributed over the group (the strong coupling limit, $\beta \rightarrow 0$). Disregarding the contributions from multiple wrappings in compact groups, we approximate the transition kernel $p_{0t}$ by the probability $p(\eta)$ of a particular noise field $\eta$,
\begin{align}
    \label{eq:gauge_transition_probability}
    p_{0t}(\mathbf{U}^{(t)} | \mathbf{U}^{(0)}) \approx p(\eta) \propto \exp[- \frac{1}{2} \sum_{a,x,\mu} (\eta^a_{x,\mu})^2].
\end{align}
It follows that the loss function is  (see \hyperref[sec:endmatter]{End Matter})
\begin{equation}
    \label{eq:adapted_gauge_loss_function}
    \! \! \int_0^T  \! { \dd t} 
    \underset{ \substack{ p_0(\mathbf{U}^{(0)}) \\ p_{0t}(\mathbf{U}^{(t)}|\mathbf{U}^{(0)}) }}
    {\mathbb{E}} 
    \left[
       \sum_{x,\mu,a} \left( s_{\theta;x,\mu}^a(\mathbf{U}^{(t)}, t) \! + \! \frac{\eta^a_{x,\mu}}{\sigma(t)} \right)^2\right].
\end{equation}
With data at $t=0$ provided by traditional MC methods, our DMs can be trained efficiently. Evidently, the score function must accurately reflect the symmetries of the theory, which are local and discrete lattice symmetries (translations, rotations, reflections). Thus, we model the score function using \mbox{L-CNNs}. We use a two-layer network which retains exact gauge equivariance and translation invariance. Since L-CNNs can learn arbitrary Wilson loops  \cite{Favoni:2020reg}, the expressiveness of the network is ensured\footnote{We note that \mbox{L-CNNs} can also be adapted to larger global symmetry groups \cite{Aronsson:2023rli}, but for computational efficiency we restrict our networks to translational symmetry.}. The time dependence of the score function is modeled by random Fourier features \cite{tancik2020fourier} (see \hyperref[sec:endmatter]{End Matter}).

\textit{Metropolis-adjusted annealing Langevin algorithm.}
Once the network is trained via the gauge invariant score matching objective of Eq.~\eqref{eq:adapted_gauge_loss_function}, the standard procedure of generating field configurations is to solve the reverse process \eqref{eq:reverse_gauge}. Here, we depart from this approach and instead apply MAALA \cite{Zhu:2024kiu, Zhu:2025pmw}, an extension of the Metropolis-adjusted Langevin algorithm \cite{roberts1998optimal, girolami2011riemann}. The key advantage of MAALA is the ability to rescale the score function of the learned probability distribution. After training at $\beta_0$, our models can be applied to different target couplings $\beta$ through \textit{score rescaling}. Additionally, Metropolis adjustments applied near the end of the generative process remove the bias of the DM. 

Using MAALA introduces a second time coordinate $\tau$, which is used to define auxiliary Langevin trajectories for each $t$, see Fig.~\ref{fig:denoise}. To perform the denoising procedure, uniformly distributed links are initialized at $t=T$ according to the Haar measure. At fixed diffusion time $t$, we then solve the discretized SDE (here, $\hat s_\theta=\sum_a T^a s^a_\theta$, $\hat\eta=\sum_a T^a \eta^a$ and we omit $x, \mu$ subscripts for brevity)
\begin{equation}
\label{eq:maala_gauge_sde}
U^{(t)}_{(\tau+\epsilon)}=
\exp\!\left[ i \left( \epsilon \frac{\beta}{\beta_0} \hat s_{\theta}(\mathbf{U}^{(t)}_{(\tau)}, t)
+\sqrt{2\epsilon} \, \hat \eta \right) \right]\, U^{(t)}_{(\tau)},
\end{equation}
where $\tau$ increases from $\tau = 0$ up to $\tau = \mathcal T$ in steps of $\epsilon$, and ${\mathcal T}$ is chosen large enough to allow for thermalization. Then, we decrease $t -\Delta t \leftarrow t$ with time step $\Delta t$ and set $\mathbf{U}^{(t-\Delta t)}_{(0)} = \mathbf{U}^{(t)}_{(\mathcal T)}$, see again Fig.~\ref{fig:denoise}. 
Eq.~\eqref{eq:maala_gauge_sde} resembles a stochastic quantization update with the time-dependent network, $s^a_{\theta;x,\mu}(\mathbf U, t)$, in place of the usual drift. The noise scale $\sigma(t)$ is not used explicitly in the update, as all information about the diffusion process is already encoded in the learned network.

Once the reverse process is close to $t=0$, we apply \textit{Metropolis adjustment} after every step. Assume we start with $\mathbf{U}^{(t)}_{(\tau)}$, and a single step of Eq.~\eqref{eq:maala_gauge_sde} yields a proposal configuration $\tilde{\mathbf{U}}^{(t)}_{(\tau)}$, we then set $\mathbf{U}^{(t)}_{(\tau+\epsilon)}=\tilde{\mathbf{U}}^{(t)}_{(\tau)}$ with probability
\begin{equation} \label{eq:p_accept}
    p_\text{accept} = \min \left\{ 1, \frac{p(\tilde{\mathbf{U}}^{(t)}_{(\tau)})}{p(\mathbf{U}^{(t)}_{(\tau)})} 
    \frac{q( \mathbf{U}^{(t)}_{(\tau)} | \tilde{\mathbf{U}}^{(t)}_{(\tau)}) }{q(\tilde{\mathbf{U}}^{(t)}_{(\tau)} | \mathbf{U}^{(t)}_{(\tau)}) } \right\},
\end{equation}
and $\mathbf{U}^{(t)}_{(\tau+\epsilon)}=\mathbf{U}^{(t)}_{(\tau)}$ otherwise. Here $p$ is determined by the target action, i.e., the Wilson action with coupling $\beta$, and the transition probabilities $q$ are determined from Eq.~\eqref{eq:maala_gauge_sde} (see \hyperref[sec:endmatter]{End Matter}). This procedure is repeated until we reach $t=0$. The adjustment step guarantees that (asymptotically with sufficiently many steps) the generated samples will be consistent with the target distribution at $\beta$.

\begin{figure}[tbp]
    \centering
    \includegraphics[width=.95\linewidth]{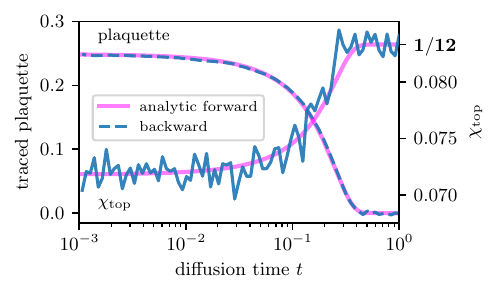}
    \caption{Comparison of the learned generative (backward) process with the forward diffusion process defined in Eq.~\eqref{eq:diffused_link} for $\beta=2$ on a $L=16$ lattice. Close agreement (up to statistical precision) between the DM and the analytical diffusion process indicates that our equivariant DMs can successfully perform denoising. At $t=1$, the forward process leads to the strong-coupling limit where the plaquette vanishes and $\chi_\mathrm{top} = 1/12$.
    }
    \label{fig:forward_backward}
\end{figure}

\begin{figure}[tbp]
    \centering
    \includegraphics[width=.95\linewidth]{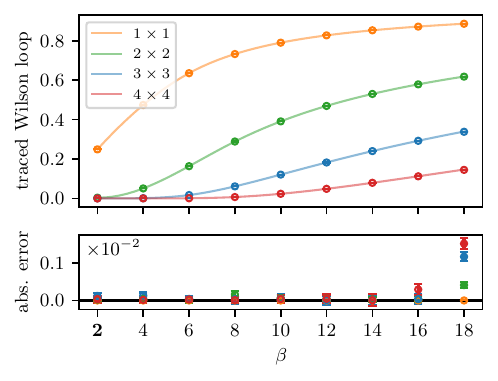}
    \includegraphics[width=.95\linewidth]{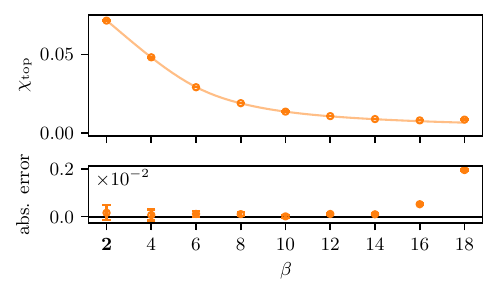}
    \caption{Predictions for two-dimensional $U(2)$ gauge theory on an $L=16$ lattice using a DM trained at $\beta=2$ (indicated in bold), $L=16$. The top panel shows predictions for $n \times n$ Wilson loops (circles with error bars) compared to their analytic values (solid lines) as a function of the inverse coupling $\beta$. The lower panel shows the topological susceptibility $\chi_\mathrm{top}$. 
    Absolute errors are presented underneath. Significant deviations only occur for the largest $\beta$ values.
    }
    \label{fig:loops}
\end{figure}

\textit{Numerical results.}
We consider a two-dimensional U(2) lattice gauge theory on a $L=16$ lattice, where we train using $100\,000$  samples generated at $\beta_0=2$ using Hybrid Monte Carlo (HMC). We train ten randomly initialized models using the Adam optimizer \cite{kingma2017adammethodstochasticoptimization}  over $1\,000$ epochs with a batch size of 64 and an exponentially decreasing learning rate starting at $10^{-5}$. Following that, $10\,000$ configurations are sampled at a range of $\beta$ values using MAALA with score rescaling. Results from different models are combined using DerSimonian-Laird analysis (see \hyperref[sec:endmatter]{End Matter}). The specific observables we examine are the expectation values of traced square Wilson loops of different sizes and the topological susceptibility $\chi_{top}$, an observable notoriously difficult to obtain using traditional MC methods. The latter is given by
\begin{align}
    \chi_\mathrm{top} = \frac{\langle Q^2 \rangle}{L^2}, \qquad Q = \frac{1}{2\pi} \sum_x \arg (\det P_{x,01}) \in \mathbb Z,
\end{align}
where $Q$ denotes the topological charge operator in the two-dimensional theory. 

To validate our results and numerical implementation, we have compared them with the analytical solution for the forward process of Eq.~\eqref{eq:diffused_link}, which can be obtained by combining the evolution determined by the Laplace-Beltrami operator in the heat kernel equation with a character expansion \cite{Bonati:2019ylr, Bonati:2019olo, Rouenhoff:2024geh}. While detailed results will be shown elsewhere, in essence coefficients in the character expansion are multiplied by $\exp(-C_2(r)\sigma^2(t))$, where $\sigma^2(t)$ is the scaled diffusion time incorporating the noise scale $\sigma(t)$, and $C_2(r)$ is the quadratic Casimir for representation $r$. Hence at the end of the forward process, all but the trivial representation are exponentially suppressed and the theory flows to its strong coupling limit, i.e.~a uniform distribution, which determines all observables. We note that this differs from the evolution on flat manifolds, such as for scalar fields, where in the variance-expanding scheme higher-order cumulants are conserved \cite{Aarts:2024rsl}. In Fig.~\ref{fig:forward_backward}, we compare the backward evolution for the plaquette and the topological susceptibility with the analytical evolution in the case of U(2) (note that the topological susceptibility equals $1/12$ in the strong-coupling limit). Excellent agreement is observed. 

Figure \ref{fig:loops} shows the out-of-distribution sampling performance of our DMs for $L=16$. We find very good agreement between model predictions and exact results for a large range of couplings up to $\beta \approx 14$. Similar results are also obtained on lattice sizes $L=32$ and $64$ (see Table~\ref{tab:larger_lattices} in the \hyperref[sec:endmatter]{End Matter}). We stress that our DMs are trained exclusively on a single ensemble  at $\beta = 2$ and $L = 16$; all other results are extrapolations using the rescaling factor in the MAALA approach. Only at the largest inverse couplings, $\beta \geq 16$, significant deviations are found.

\begin{figure}[tbp]
    \centering
    \includegraphics[width=.95\linewidth]{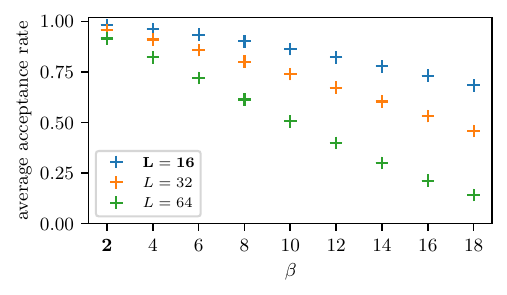}
    \caption{Average Metropolis acceptance rates for extrapolations to different $\beta$ and lattice sizes. DMs are trained at $\beta=2$ and $L=16$.
    }
    \label{fig:acceptance_rate}
\end{figure}

Figure \ref{fig:acceptance_rate} shows the measured acceptance rates of the Metropolis step, as defined by Eq.~\eqref{eq:p_accept}. We use the acceptance rate as a measure of the mismatch between the learned score of the DM and the true action. At the training point $\beta=2$, we observe only a slight deterioration with increased lattice size. For $L=16$ we find only a moderate decrease for larger $\beta$. Indeed, only the combination of large $\beta$ and large $L$ leads to a significant deterioration of the sample quality. At this point the mismatch between the reweighted score function and the true action is too large to be effectively corrected by the Metropolis adjustment, which affects the predictions for observables as well. We stress that the measured acceptance rates appear to have less-than-exponential scaling with the lattice volume. 

\begin{figure}[tbp]
    \centering
    \includegraphics[width=.95\linewidth]{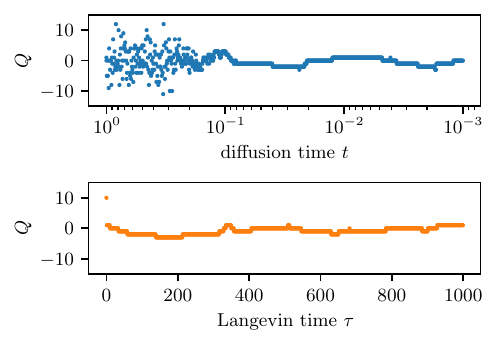}
    \caption{Evolution of the topological charge during the generative process of one field configuration for $\beta=14$ on a $L=16$ lattice. The top panel shows the evolution for the MAALA along (reverse) diffusion time $t$. In the bottom panel, the evolution along a Langevin trajectory using stochastic quantization with the Wilson action is shown. To enable a fair comparison, we use the same total number of Langevin steps and the same step size for both processes. 
    While the Langevin trajectory always remains trapped in a few topological sectors for extended periods, MAALA initially exhibits frequent transitions.
    }
    \label{fig:topological_freezing}
\end{figure}

In Fig.~\ref{fig:topological_freezing} we illustrate the advantage of the annealing in MAALA in comparison to stochastic quantization, where one solves Eq.~\eqref{eq:maala_gauge_sde} with the gradient of the Wilson action in place of the network. While SQ exhibits dramatic topological freezing at $\beta=14$, the reverse process using MAALA allows for rapid exploration of the topological sectors during its initial annealing phase.

Finally, we have also applied our approach to two-dimensional SU(2) gauge theory. Performance is comparable to, if not better than, the U(2) case, see Table~\ref{tab:su2_results} in the \hyperref[sec:endmatter]{End Matter}.

\textit{Conclusions and outlook.} In this Letter, we have demonstrated that gauge equivariant DMs can learn to accurately sample non-Abelian gauge theories on the lattice. Through score rescaling, our models (trained on a single MC ensemble) can access a broad range of couplings and lattice sizes. Moreover, the generated samples are independent, which allows us to effectively circumvent critical slowing down and topological freezing. The remaining bias of our DMs can be corrected through Metropolis adjustment up to high inverse couplings. We have found that it is essential to use network architectures which respect gauge symmetry. By using L-CNNs, our approach is fully compatible with the relevant symmetries, extends to other gauge groups and can be formulated in arbitrary space-time dimensions.

The question remains if DMs can successfully scale to four-dimensional SU(3) gauge theory with large lattice volumes. The relatively small loss of acceptance rate for larger lattices at the training point certainly prompts further research in this direction. At larger couplings, the scaling appears to be weaker than the exponential scaling of normalizing flow approaches \cite{Abbott:2022zsh}. Even if it turns out that large volumes remain challenging for DMs, then another particularly promising application is to sample ensembles based on fixed-point actions where highly accurate but computationally costly approximations have been found using \mbox{L-CNNs} \cite{Holland:2024muu, Holland:2025fsa}. The fixed-point approach does not require large lattice volumes as lattice artifacts are strongly suppressed by design and DMs could provide substantial speed-ups to already existing HMC simulations. Importantly, generating training ensembles with expensive actions only has to happen once, after which the DM takes over to generalize.
Furthermore, DMs could be extended to include fermionic fields similar to flow-based models \cite{Albergo:2021bna, Abbott:2022zhs}.

\begin{acknowledgments} 
\textit{Acknowledgments.} 
We thank Biagio Lucini and Kai Zhou for insightful discussions.
GA thanks LW for hospitality at RIKEN and the University of Tokyo during the completion of this paper.
% Institutions
We thank the DEEP-IN working group at RIKEN iTHEMS for support in the preparation of this paper.
% Funds
GA is supported by STFC grant ST/X000648/1 and by a Royal Society Leverhulme Trust Senior Research Fellowship. 
DEH is supported by the UKRI AIMLAC CDT EP/S023992/1. 
AI, DM and TR acknowledge funding from the Austrian Science Fund (FWF) project PAT3667424.
LW is supported by JSPS KAKENHI Grant No.~25H01560, and JST-BOOST Grant No.~JPMJBY24H9. LW and QZ are also supported by the RIKEN-TRIP initiative (RIKEN-Quantum), 
WW and QZ are supported in part by the Natural Science Foundation of China under grants No.~12125503 and 12305103.  
The computational results have been achieved using the Austrian Scientific Computing (ASC) infrastructure.
\end{acknowledgments}

\textit{Research Data and Code Access} -- The code and data used for this manuscript will be available in version 2 on the arXiv.

\textit{Open Access Statement} – For the purpose of open access, the authors have applied a Creative Commons Attribution (CC BY) licence to any Author Accepted Manuscript version arising.

\bibliography{refs_letter}

\clearpage

\section{End Matter}
\label{sec:endmatter}

\textit{Group derivatives.} 
Derivatives of a function $f$ on the group manifold $G$ are defined via
\begin{align}
     D^a f(U) = \lim_{\epsilon \rightarrow 0} \frac{1}{\epsilon} (f(e^{i \epsilon  T^a} U) - f(U) ) = \frac{\mathrm{d}}{\mathrm{d}\epsilon}f(e^{i\epsilon  T^a}U) \big\rvert_{\epsilon=0}.
\end{align}
Intuitively, the group derivative measures how the function responds to a perturbation \emph{parallel} to the group manifold. For functions which depend on a lattice of links, the derivative depends additionally on the indices $x,\mu$ which denote the perturbed link.

To efficiently train DMs using the loss function in Eq.~\eqref{eq:gauge_loss_function}, we need to calculate the group derivative of $\ln(p_{0t}(\mathbf{U}^{(t)} | \mathbf{U}^{(0)}))$ with respect to the first argument. Following Eq.~\eqref{eq:gauge_transition_probability}, we have to calculate
\begin{align}
     D^a_{x,\mu} \ln p_{0t}  =  -D^a_{x,\mu} \mathrm{Tr}(\hat \eta^2_{x,\mu}),
\end{align}
where we have used $\hat \eta = \sum_a T^a \eta^a$ and  $\frac{1}{2}\sum_{a} (\eta^a)^2 = \mathrm{Tr}(\hat \eta^2)$. For brevity, we drop the $x$ and $\mu$ indices. Using Eq.~\eqref{eq:diffused_link}, we express $\hat \eta$ as a matrix logarithm
\begin{align}
    \hat \eta = -\frac{i}{\sigma(t)} \ln \left( U^{(t)} U^{(0) \dagger} \right).
\end{align}
Thus, we can write
\begin{align} \label{eq:transition_kernel_derivative}
     D^a \ln p_{0t} &= \frac{1}{\sigma^2(t)} \mathrm{Tr} \left[ \frac{\mathrm{d}}{\mathrm{d} \epsilon} \ln \left( e^{i\epsilon T^a} U^{(t)} U^{(0) \dagger} \right)^2 \right]_{\epsilon=0} \nonumber \\
     &= \frac{2i}{\sigma^2(t)} \mathrm{Tr} \left[ T^a \ln \left( U^{(t)} U^{(0) \dagger} \right) \right], \nonumber \\
     &= -\frac{1}{\sigma(t)}\eta^a_{x,\mu},
\end{align}
where we have used the Baker–Campbell–Hausdorff formula and \mbox{$\mathrm{Tr} ([A,B]B)=0$}.
Inserting Eq.~\eqref{eq:transition_kernel_derivative} into Eq.~\eqref{eq:gauge_loss_function} yields the adapted loss function in Eq.~\eqref{eq:adapted_gauge_loss_function}.

\textit{Training and model details.} 
For training, the stochastic process of Eq.~\eqref{eq:diffused_link} is solved on a logarithmic diffusion time grid consisting of $N_t = 800$ points from $t=0$ to $t= T \equiv 1$ such that more steps are performed closer to $t=0$. Our models for the score function consist of two \mbox{L-CNN} bilinear convolution layers with kernel size $K=2$ with a single input channel for the plaquette $P_{x,01}$ and two output channels corresponding to the two lattice directions. The hidden layer consists of 16 channels. We define the bilinear layer as
\begin{align}
    \hat W'^{l}_{x} = \sum_{k, \mu, m,n} \omega^{lmn}_{k,\mu} \,  \hat W^m_{x} \, U_{x, k \hat \mu} \hat W^n_{x+ k \hat \mu}  U^\dagger_{x, k \hat \mu},
\end{align}
where $\omega$ are the trainable parameters and the size of the kernel is constrained by \mbox{$-(K-1) \leq k \leq K-1$}. The index $l$ denotes the output channel and $m$ and $n$ are summed over the input channels. The matrices $\hat{W}$ are locally transforming variables, which are the plaquettes in the input layer. The matrices $U_{x, k \hat \mu}$ denote the ordered product of links starting at $x$ and ending at $x + k \cdot \hat \mu$, which forms a straight Wilson line. In each layer, the input channels are doubled by including the Hermitian conjugates, which is necessary to represent all orientations of a loop. By additionally including unit matrices as part of the input channels, the bilinear convolutional layer can also represent invariant bias terms and linear convolutions. For the model architecture used in this work, the number of trainable parameters for the bilinear convolution amounts to 528 for the first and 10\,626 for the second layer.

To incorporate the time dependence of the score function, we use random Fourier features \cite{tancik2020fourier} with embedding dimension 256, which includes a single fully-connected dense layer with input size 256 and an output dimension of 16. The number of trainable parameters of this layer is 4112. The resulting embedding vector $f^l(t)$ is added to the locally transforming variables between the two bilinear convolution layers by
\begin{align}
    \hat W'^{l}_{x} = \hat W^{l}_{x} + \mathds{1} \, f^l(t),
\end{align}
which retains gauge equivariance of the network. For consistency, the output dimension of the time embedding must match the size of the hidden L-CNN layer. In total our architecture has 15\,266 trainable parameters.

\textit{MAALA implementation details.}
The transition probabilities $q( \mathbf{U}^{(\tau)} | \tilde{\mathbf{U}}^{(\tau)}) $ and $q(\tilde{\mathbf{U}}^{(\tau)} | \mathbf{U}^{(\tau)})$ needed for the Metropolis adjustment are given by the probabilities of the noise $\eta $ required for the update steps $\mathbf{U}^{(\tau)} \mapsto \tilde{\mathbf{U}}^{(\tau)}$ and $\tilde{\mathbf{U}}^{(\tau)} \mapsto \mathbf{U}^{(\tau)}$.
The noise can be extracted by rewriting the discretized SDE \eqref{eq:maala_gauge_sde}, and using the abbreviations $\hat s_\theta = \sum_{a} T^a s^a_{\theta}$, $\hat \eta  = \sum_{a} T^a \eta^{a}$, as
\begin{align}
    \hat \eta_{x,\mu} &= \frac{1}{i \sqrt{2 \epsilon}} \ln(\tilde{{U}}^{(\tau)}_{x,\mu} U_{x,\mu}^{(\tau) \dagger}) - \sqrt{\frac{\epsilon}{2}} \hat s_{\theta;x,\mu}(\mathbf{U}^{(\tau)}), \\
    \hat \eta'_{x,\mu} &= \frac{1}{i \sqrt{2 \epsilon}} \ln( U^{(\tau)}_{x,\mu} \tilde{U}^{(\tau)\dagger}_{x,\mu}) - \sqrt{\frac{\epsilon}{2}} \hat s_{\theta;x,\mu} (\tilde{\mathbf{U}}^{(\tau)}).
\end{align}
This leads to an Metropolis acceptance rate of
\begin{align}
    \begin{split}
    \min \big\{ 1, \exp \big( 
    &-(S_E(\tilde{\mathbf{U}}^{(\tau)}) - S_E(\mathbf{U}^{(\tau)})) \\
    &+ \sum_{x,\mu} \big[ \mathrm{Tr}[\hat \eta^2_{x,\mu}]  - \mathrm{Tr}[\hat \eta'^2_{x,\mu}] \big]
    \big) \big\}.
    \end{split}
\end{align}

For sample generation using MAALA, we use the same logarithmically spaced diffusion time steps  as during training. At each diffusion time step, we perform $N_\tau = 250$ Langevin update steps with time step $\epsilon = 5 \cdot 10^{-3}$, before decreasing $t$ to the next time step. In the last $2 \%$ of time steps, Metropolis adjustment is performed. We note that we do not adapt these parameters when changing the inverse coupling at which we sample. For small $\beta$, significantly less update steps may be sufficient. Generating $10\,000$ field configurations at $L=16$ takes around 20 hours on a single NVIDIA H100 SXM5 GPU. 

\begin{table}[!htb]
\caption{Numerical results for U(2) for larger Wilson loops and the topological susceptibility at $L = 16$, $32$ and $64$ and three separate inverse couplings $\beta$. The DM was trained on $\beta=2$, $L=16$ (highlighted). For $L=32$ and $L=64$ only $1000$ samples were generated.\label{tab:larger_lattices}}
\begin{ruledtabular}
\scriptsize
\begin{tabular}{l| llll}
 $\bm{\beta=2}$ & $\;\;\;2 \times 2$ & $\;\;\;3 \times 3$ & $\;\;\;4 \times 4$ & $\;\;\;\chi_\mathrm{top}$ \\
$\bm{L=16}$ & 0.00381(7) &  $1.3(7) \cdot 10^{-4}$   & $-3(9) \cdot 10^{-5}$ & 0.0716(3) \\
$L=32$ & 0.0038(1) & $-1(1) \cdot 10^{-4}$  & $0(1) \cdot 10^{-4}$ & 0.070(1)\\ 
$L=64$ & 0.00378(6) & $2(6) \cdot 10^{-5}$ & $5(7) \cdot 10^{-5}$ & 0.072(1)  \\
exact & 0.003835 & $3.660 \cdot 10^{-6}$ & $2.163 \cdot 10^{-10}$ & 0.071802 \\
\hline
$\beta=10$ & &  &  &   \\
$L=16$ & 0.39106(9) & 0.1209(1)  & 0.0234(1) & 0.01345(8) \\
$L=32$ & 0.3909(2) & 0.1210(2)  & 0.0236(2) & 0.0137(2) \\
$L=64$ & 0.39108(7) & 0.12102(9) & 0.02335(8) & 0.0135(2)  \\
exact & 0.390989 & 0.120884 & 0.023370 & 0.013442 \\
\hline
$\beta=18$ & &  &  &   \\
$L=16$ & 0.61824(7) & 0.3383(1)  & 0.1450(1) & 0.00837(5) \\
$L=32$ & 0.6184(1) & 0.3390(2)  & 0.1459(2) & 0.0084(1) \\
$L=64$ & 0.6184(1) & 0.3390(2) & 0.1460(2) & 0.0084(1)  \\
exact & 0.618647 & 0.339427 & 0.146478 & 0.006399 \\
\end{tabular}
\end{ruledtabular}
\end{table}

\begin{table}[!htb]
\caption{Numerical results for SU(2) calculations on $L=16$, $32$ and $64$ lattices for Wilson loops at several inverse couplings $\beta$. The DM was trained on $\beta=2$, $L=16$.
\label{tab:su2_results}}
\begin{ruledtabular}
\scriptsize
\begin{tabular}{l| llll}
 $\bm{\beta=2}$ & $\;\;\;1 \times 1$ & $\;\;\;2 \times 2$ & $\;\;\;3 \times 3$ & $\;\;\;4 \times 4$ \\
$\bm{L=16}$ & 0.4333(2) & 0.0357(3) & 0.0006(3) & $-0.0003(3)$ \\
$L=32$ & 0.4333(4) & 0.0349(5) & 0.0014(5) & $-0.0001(5)$ \\
$L=64$ & 0.4333(2) & 0.0351(3) & 0.0002(3) & 0.0004(2) \\
exact & 0.433127 & 0.035194 & 0.000536 & $1.534 \cdot 10^{-6}$ \\
\hline
$\beta=10$ & &  &  &   \\
$L=16$ & 0.85420(7) & 0.5325(3) & 0.2421(5) & 0.0807(5) \\
$L=32$ & 0.8541(1) & 0.5326(4) & 0.2422(8) & 0.0808(8) \\
$L=64$ & 0.85420(5) & 0.5323(2) & 0.2421(4) & 0.0805(4) \\
exact & 0.854185 & 0.532364 & 0.242086 & 0.080322 \\
\hline
$\beta=18$ & &  &  &   \\
$L=16$ & 0.91793(4) & 0.7097(2) & 0.4621(4) & 0.2533(6) \\
$L=32$ & 0.91782(7) & 0.7096(3) & 0.4620(7) & 0.253(1) \\
$L=64$ & 0.91792(3) & 0.7099(2) & 0.4628(4) & 0.2545(5) \\
exact & 0.917894 &  0.709854 & 0.462520 & 0.253909 \\
\end{tabular}
\end{ruledtabular}
\end{table}

\textit{Detailed numerical results.}
We have tested our DMs trained on $\beta=2$, $L=16$ for U(2) gauge theory across a wide range of inverse couplings and lattice sizes. The results are summarized in Table~\ref{tab:larger_lattices}. We observe only negligible dependence of the observables on the lattice size. All predictions except for $\beta=18$ are compatible with the analytic calculation within statistical precision. Deviations occur for the largest loops and the topological susceptibility, which are observables sensitive to the infrared modes of the theory. 

Results for SU(2) are summarized in Table~\ref{tab:su2_results}, which shows even better agreement at $\beta = 18$. Here, we only investigate the square Wilson loops, as the topological charge is always zero in two-dimensional SU(2) theory. 

\textit{Error estimates.} 
We combine the predictions of our independently trained networks using the DerSimonian-Laird analysis \cite{DerSimonian:1986aaa, DerSimonian:2015aaa, Cochran:1954aaa} (see also Ref.~\cite{Aarts:2025lpi}).
Given a set of measured mean values $O_i$ and uncertainty estimates $\delta O_i$, where the index $i=1,...,N$ (with $N=10$) enumerates the trained networks, we calculate the weighted mean as
\begin{align}
    \bar{O} = \frac{\sum_i w_i O_i}{\sum_i w_i}, \quad w_i=\frac{1}{\delta O_i^2}.
\end{align}
Then, we compute the quantities 
\begin{align}
    \tau^2 &= \max(0, \frac{Q-(N-1)}{C}), \\
    Q &= \sum_i w_i (O_i - \bar{O})^2, \\ 
    C &= \sum_i w_i - \frac{\sum_i w_i^2}{\sum_i w_i},
\end{align}
to define the adjusted weights
\begin{equation}
    w^*_i = \frac{1}{\delta O_i^2 + \tau^2},
\end{equation}
and the adjusted weighted mean and the total random-effects uncertainty
\begin{align}
    O^* = \frac{\sum_i w^*_i O_i}{\sum_i w^*_i}, \quad \sigma^2_\mathrm{total} = \frac{1}{\sum_i w^*_i}.
\end{align}

\end{document}